\documentclass[aps,prc,showpacs,twocolumn,amssymb,superscriptaddress,floatfix]{revtex4}

\usepackage[colorlinks,citecolor=blue]{hyperref}
\usepackage{graphicx}
\usepackage{amsmath, amsthm, amssymb}

\def\be{\begin{eqnarray}}
\def\ee{\end{eqnarray}}
\def\lsim{\stackrel{\scriptstyle <}{\phantom{}_{\sim}}}
\def\gsim{\stackrel{\scriptstyle >}{\phantom{}_{\sim}}}

\begin{document}
\title{Rise of azimuthal anisotropies as a signature of the Quark-Gluon-Plasma\\
in relativistic heavy-ion collisions}

\author{V. P. Konchakovski}
\affiliation{Institute for Theoretical Physics, University of Giessen, Giessen, Germany}
\affiliation{Bogolyubov Institute for Theoretical Physics, Kiev, Ukraine}

\author{E. L. Bratkovskaya}
\affiliation{Institute for Theoretical Physics, University of Frankfurt, Frankfurt, Germany}
\affiliation{Frankfurt Institute for Advanced Studies, Frankfurt, Germany}

\author{W. Cassing}
\affiliation{Institute for Theoretical Physics, University of Giessen, Giessen, Germany}

\author{V. D Toneev}
\affiliation{Frankfurt Institute for Advanced Studies, Frankfurt, Germany}
\affiliation{Joint Institute for Nuclear Research, Dubna, Russia}

\author{V. Voronyuk}
\affiliation{Bogolyubov Institute for Theoretical Physics, Kiev, Ukraine}
\affiliation{Frankfurt Institute for Advanced Studies, Frankfurt, Germany}
\affiliation{Joint Institute for Nuclear Research, Dubna, Russia}

\begin{abstract}
The azimuthal anisotropies of the collective transverse flow of
hadrons are investigated in a large range of heavy-ion collision
energy within the Parton-Hadron-String Dynamics (PHSD) microscopic
transport approach which incorporates explicit partonic degrees of
freedom in terms of strongly interacting quasiparticles (quarks
and gluons) in line with an equation-of-state from lattice QCD as
well as dynamical hadronization and hadronic dynamics in the final
reaction phase. The experimentally observed increase of the
elliptic flow $v_2$ with bombarding energy is successfully
described in terms of the PHSD approach in contrast to a variety
of other kinetic models based on hadronic interactions. The
analysis of higher-order harmonics $v_3$ and $v_4$ shows a similar
tendency of growing deviations between partonic and purely
hadronic models with increasing bombarding energy. This signals
that the excitation functions of azimuthal anisotropies provide a
sensitive probe for the underling degrees of freedom excited in
heavy-ion collisions.
\end{abstract}


\maketitle

{\it Introduction.}\
A few decades of experimental studies at the Schwerionen-Synchrotron
(SIS), the Alternating Gradient Synchroton (AGS) and the Super
Proton Synchroton (SPS) have shown that the physics of nuclear
collisions at moderate relativistic energies is dominated by the
nonequilibrium dynamics of hadronic resonance matter, {\it i.e.} the
confined phase of QCD. The body of data extends and builds up the
knowledge gained about dense hadronic matter, in particular, at the
SPS/CERN. The SPS heavy-ion data have shown several signatures
that hinted at the onset of a quark-gluon plasma (QGP)
formation~\cite{HJ00, Gazd}. With the
Relativistic Heavy Ion Collider (RHIC) the center-of-mass energy
could be increased by a factor of 10 relative to the SPS, and the
experiments at RHIC assured that a new form of matter -- well above
the deconfinement transition point -- was created in the
laboratory.

Indeed, the discovery of a large azimuthal anisotropic flow of
hadrons at RHIC provides a conclusive evidence for the creation of
dense partonic matter in ultra-relativistic nucleus-nucleus
collisions. The strongly interacting medium in the collision zone
can be expected to achieve a local equilibrium and exhibit an
approximately hydrodynamic flow~\cite{Ol92,HK02,Sh09}. The momentum
anisotropy is generated due to pressure gradients in a collective
expansion of an initial geometry of an ``almond-shaped'' collision
zone produced in noncentral collisions~\cite{Ol92,HK02}. The
pressure gradients translate early stage coordinate space asymmetry
to final-state momentum space anisotropy~\cite{Pet123}. The
picture thus emerges that the medium created in ultra-relativistic
collisions for a couple of fm/$c$ interacts more strongly than hadron
resonance matter and exhibits collective properties that resemble
those of a liquid of a very low shear viscosity $\eta$ to the
entropy density $s$ ratio, $\eta/s$, close to a nearly perfect
fluid~\cite{Sh05,GMcL05,PC05}.

An experimental manifestation of this collective flow is the
anisotropic emission of particles in the plane transverse to the
beam direction. A quark number scaling of the elliptic flow
proposed in Ref.~\cite{Vol02} was observed at RHIC for a broad range of
particle species, collision centralities, and transverse kinetic
energy, presumably interpreted as due to the development of
substantial collectivity in the early partonic phase~\cite{Lac07}.

It was shown that higher-order anisotropy harmonics, in particular
the hexadecupole moment $v_4$, can provide a more sensitive
constraint on the magnitude of $\eta/s$ and the freeze-out dynamics,
and the ratio $v_4/(v_2)^2$ might indicate whether a full local
equilibrium is achieved in the QGP~\cite{Bh05}. Recently, the
importance of the triangular flow $v_3$, which originates from
fluctuations in the initial collision geometry, has been pointed
out~\cite{AR10,XK10}. The participant triangularity characterizes
the triangular anisotropy of the initial nuclear overlap geometry
and arises from event-by-event fluctuations in the
participant-nucleon collision points and corresponds to a large
third Fourier component in two-particle azimuthal correlations at
large pseudo-rapidity separation $\Delta\eta$. This fact suggests a
significant contribution of the triangular flow to the ridge
phenomenon and broad away-side structures observed in the RHIC
data~\cite{AR10}.

A large number of anisotropic flow measurements have been performed
by many experimental groups at SIS, AGS, SPS and RHIC energies over
the past 20 years. Very recently the azimuthal asymmetry has
 also been measured at the Large Hadron Collider (LHC) at CERN~\cite{LHC}. 
The Beam Energy Scan (BES) program proposed at
RHIC~\cite{ST11} covers the energy interval from $\sqrt{s_{NN}}=$
200~GeV, where partonic degrees of freedom (DOF) play a decisive role,
down to the AGS energy $\sqrt{s_{NN}}\approx$ 5~GeV, where most
experimental data can be described successfully in terms of hadronic
DOF Lowering the collision energy and studying
the energy dependence of an anisotropic flow allows one to search for
the onset of the transition to a phase with partonic
DOF at an early stage of the collision as well as
possibly identify the location of the expected critical end-point
that terminates the first order phase transition at high
quark-chemical potential~\cite{Lac07,Agg07}.

This work aims to study excitation functions for different harmonics
of the charged particle anisotropy in momentum space in a wide
collision energy range, i.e.\ from the AGS to the top RHIC energy
regime. We want to clarify how the interplay of quark and hadron
DOF is changed with increasing bombarding energy. In this study
we investigate the excitation function of different flow
coefficients. Our analysis of the STAR/PHENIX RHIC data -- based on
recent results of the BES program -- will be performed within the
PHSD transport approach~\cite{PHSD} that includes explicit partonic
DOF as well as a dynamic hadronization scheme for the transition
from partonic to hadronic DOF and vice versa.

{\it The PHSD approach.}\
The dynamics of partons, hadrons and strings in relativistic
nucleus-nucleus collisions is analyzed here within the
Parton-Hadron-String Dynamics approach~\cite{PHSD}. In this
transport approach the partonic dynamics is based on Kadanoff-Baym
equations for Green functions with self-energies from the Dynamical
QuasiParticle Model (DQPM)~\cite{Cassing06,Cassing07} which
describes QCD properties in terms of ``resummed'' single-particle
Green functions. In Ref.~\cite{BCKL11}, the actual three DQPM
parameters for the temperature-dependent effective coupling were
fitted to the recent lattice QCD results of Ref.~\cite{aori10}.
The latter lead to a critical temperature $T_c \approx$ 160~MeV
which corresponds to a critical energy density of $\epsilon_c
\approx$ 0.5~GeV/fm$^3$. In PHSD the parton spectral functions
$\rho_j$ ($j=q, {\bar q}, g$) are no longer $\delta$-functions in
the invariant mass squared as in conventional cascade or transport
models but depend on the parton mass and width parameters which
were fixed by fitting the lattice QCD results from
Ref.~\cite{aori10}. We recall that the DQPM allows one to extract a potential
energy density $V_p$ from the space-like part of the energy-momentum
tensor as a function of the scalar parton density $\rho_s$.
Derivatives of $V_p$ with respect to $\rho_s$ then define a scalar mean-field
potential $U_s(\rho_s)$ which enters into the equation of motion for the
dynamic partonic quasiparticles. Furthermore, a two-body
interaction strength can be extracted from the DQPM as well from the
quasiparticle width in line with Ref.~\cite{PC05}. The transition
from partonic to hadronic DOF (and vice versa) is
described by covariant transition rates for the fusion of
quark-antiquark pairs or three quarks (antiquarks), respectively,
obeying flavor current-conservation, color neutrality as well as
energy-momentum conservation~\cite{PHSD,BCKL11}. Since the dynamical
quarks and antiquarks become very massive close to the phase
transition, the formed resonant ``prehadronic'' color-dipole states
($q\bar{q}$ or $qqq$) are of high invariant mass, too, and
sequentially decay to the ground-state meson and baryon octets
increasing the total entropy.

On the hadronic side PHSD includes explicitly the baryon octet and
decouplet, the $0^-$- and $1^-$-meson nonets as well as selected
higher resonances as in the Hadron-String-Dynamics (HSD) 
approach~\cite{Ehehalt,HSD}. Hadrons of higher
masses ($>$ 1.5~GeV in case of baryons and $>$ 1.3~GeV for
mesons) are treated as ``strings'' (color dipoles) that decay to the
known (low-mass) hadrons, according to the JETSET algorithm~\cite{JETSET}. 
Note that PHSD and HSD merge at low energy density, in
particular below the critical energy density $\epsilon_c\approx$
0.5~GeV/fm$^{3}$.

The PHSD approach was applied to nucleus-nucleus collisions from
$\sqrt{s_{NN}}\sim$ 5 to 200~GeV in Refs.~\cite{PHSD,BCKL11} in
order to explore the space-time regions of ``partonic matter''. It was
found that even central collisions at the top-SPS energy of
$\sqrt{s_{NN}}=$ 17.3~GeV show a large fraction of nonpartonic, i.e., 
hadronic or stringlike matter, which can be viewed as a
hadronic corona. This finding implies that neither hadronic nor only
partonic ``models'' can be employed to extract physical conclusions in
comparing model results with data. All these previous findings
provide promising perspectives to use PHSD in the whole range from
about $\sqrt{s_{NN}}=$ 5 to 200~GeV for a systematic study of
azimuthal asymmetries of hadrons produced in relativistic
nucleus-nucleus collisions.

\begin{figure}[t]
\centering{ \includegraphics[height=6truecm,clip]{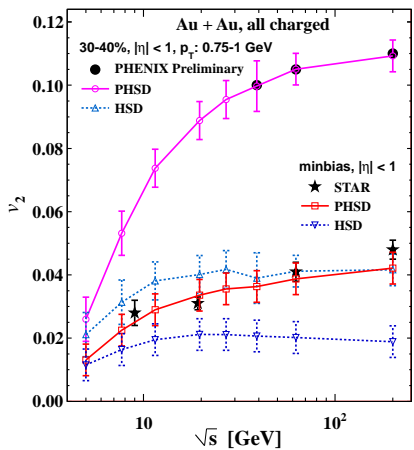}}
\caption{Average elliptic flow $v_2$ of charged particles at
  midrapidity for two centrality selections calculated within the PHSD
  (solid curves) and HSD (dashed curves) approaches. The $v_2$ STAR
  data compilation for minimal bias collisions are taken
  from Ref.~\cite{NKKNM10} (stars) and the preliminary PHENIX
  data~\cite{PHENIX_v2_s} are plotted by solid circles.} 
\label{vns}
\end{figure}

{\it Calculational results and comparison to data.}\
The anisotropy in the azimuthal angle $\psi$ is usually
characterized by the even order Fourier coefficients $v_n =\langle
exp(\, \imath \, n(\psi-\Psi_{RP}))\rangle, \
 n = 2, 4, ...$, since for a smooth angular profile the odd harmonics
become equal to zero. As noted above, $\Psi_{RP}$ is the azimuth of the
reaction plane and the brackets denote averaging over particles
and events. In particular, for the widely used second-order coefficient,
denoted as an elliptic flow, we have
\be \label{eqv2}
 v_2 = \left<cos(2\psi-2\Psi_{RP})\right>=
 \left<\frac{p^2_x - p^2_y}{p^2_x + p^2_y}\right>~,
\ee
where $p_x$ and $p_y$ are the $x$ and $y$ components of the particle
momenta. This coefficient can be considered as a function of
centrality, pseudorapidity $\eta$, and/or transverse momentum $p_T$. We
note that the reaction plane in PHSD is given by the $(x - z)$ plane
with the $z$ axis in the beam direction.
Integrated $v_3$ and $v_4$ coefficients are calculated by the
two-particle correlation method in line with Ref.~\cite{corrV2}:
\be \langle cos(n\psi_1-n\psi_2)) \rangle = \langle v_n^2 \rangle +
\delta_n \ ,\ee
which is based on the assumption that nonflow contributions
$\delta_n$ are small and account for event-by-event fluctuations of
the event plane.
In Fig.~\ref{vns} the experimental $v_2$
excitation function in the transient energy range is compared to the
results from the PHSD calculations; HSD model results are given as
well for reference. We note that the centrality selection and
acceptance are the same for the data and models.

We recall that the HSD model has been very successful in describing
heavy-ion spectra and rapidity distributions from SIS to SPS
energies. A detailed comparison of HSD results with respect to a
large experimental data set was reported in Ref.~\cite{BRAT04}
for central Au+Au (Pb+Pb) collisions from AGS to top SPS energies.
Indeed, as shown in Fig.~\ref{vns} (dashed lines), HSD is in good
agreement with experiment for both data sets at the lower edge
($\sqrt{s_{NN}}\sim$ 10~GeV) but predicts an approximately
energy-independent flow $v_2$ at larger energies and, therefore, does
not match the experimental observations. This behavior is in quite
close agreement with another independent hadronic model, the UrQMD
(Ultra relativistic Quantum Molecular Dynamics)~\cite{UrQMD} (cf.\
with~\cite{NKKNM10}).

From the above comparison one may conclude that the rise of $v_2$
with bombarding energy is not due to hadronic interactions and
models with partonic DOF have to be addressed.
Indeed, the PHSD approach incorporates the parton medium effects in
line with a lQCD equation of state, as discussed above, and also
includes a dynamic hadronization scheme based on covariant
transition rates. It is seen from Fig.~\ref{vns} that PHSD performs
better: The elliptic flow $v_2$ from PHSD (solid curve) is fairly in
line with the data from the STAR and PHENIX collaborations and
clearly shows the growth of $v_2$ with the bombarding energy.

Since partonic DOF come into play, it is interesting to compare
with another parton-hadron model, i.e. the AMPT (A Multi Phase
Transport) model~\cite{AMPT}. This model is based on a perturbative
QCD description of partonic interactions, including the production
of multiple minijet partons according to the number of binary
initial collisions. As shown in Ref.~\cite{NKKNM10}, the AMPT model
predicts an approximately constant $v_2$ with $\sqrt{s_{NN}}$
similar to the hadronic models HSD and UrQMD; however, the $v_2$
values match the experimental data at the top RHIC energy. This
discrepancy is due to a pQCD description of the partonic phase in
AMPT where the minijet partons are treated as massless and their
potentials are disregarded when they undergo scattering. Note that
PHSD and AMPT (with the additional strong hadron melting assumption
in AMPT) practically give the same elliptic flow at the top RHIC
energy of $\sqrt{s_{NN}}=$ 200~GeV. We note that the PHSD model
includes more realistic properties of dynamical quasiparticles
especially in the vicinity of the critical energy density.
Furthermore, the quark-gluon transport in PHSD naturally passes on
to the (hadronic) HSD model at lower $\sqrt{s_{NN}}$.

\begin{figure}[tb]
\centering{ \includegraphics[height=6truecm,clip]{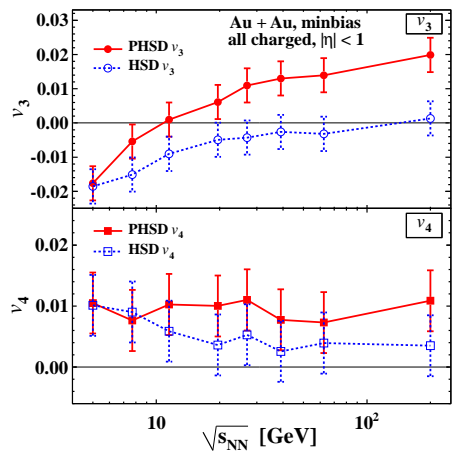}} 
\caption{Average anisotropic flows $v_3$ and $v_4$ of charged
  particles at mid-pseudorapidity for minimum bias collisions of
  $Au+Au$ calculated within the PHSD (solid lines) and HSD (dashed
  lines) models.} 
\label{vns34}
\end{figure}

In Fig.~\ref{vns34} we display the PHSD and HSD results for the
anisotropic flows $v_3$ and $v_4$ of charged particles at
midpseudorapidity for $Au+Au$ collisions from $\sqrt{s_{NN}}=$ 5 to
200~GeV. The triangular flow increases with $\sqrt{s_{NN}}$ having
negative values for $\sqrt{s_{NN}} \lsim$ 10~GeV. The pure hadronic
model HSD gives $v_3\approx$ 0 for $\sqrt{s_{NN}}\gsim$ (20-30) GeV.
Accordingly, the results from PHSD (solid red lines) are
systematically larger than those from HSD (dashed blue lines).
Unfortunately, our statistics is not high enough to allow for more
precise conclusions. The hexadecupole flow $v_4$ stays almost
constant in the considered energy range; here PHSD gives slightly
higher values than HSD.

\begin{figure}[t]
\centering{ \includegraphics[height=4.5truecm,clip]{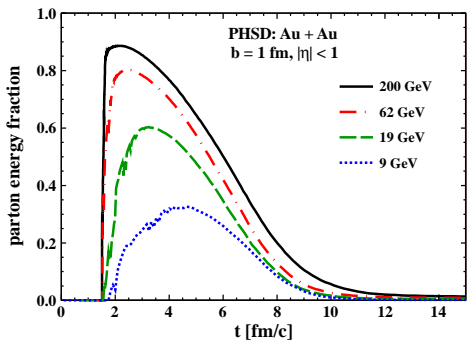} } % ,trim=0 30 0 50
\caption{ The evolution of the parton fraction of the total energy
 density at the mid-pseudorapidity for different collision energies.}
\label{part}
\end{figure}

The $v_2$ increase is clarified in Fig.~\ref{part} where the
partonic fraction of the energy density at mid-pseudorapidity with
respect to the total energy density in the same pseudorapidity
interval is shown. We recall that the repulsive scalar mean field
potential $U_s(\rho_s)$ for partons in the PHSD model leads to an
increase of the flow $v_2$ as compared to that for HSD or PHSD
calculations without partonic mean fields~\cite{BCKL11}. As follows
from Fig.~\ref{part}, the energy fraction of the partons
substantially grows with increasing bombarding energy while the
duration of the partonic phase is roughly the same.
Accordingly, the increasing influence of the repulsive partonic
mean-field $U_s(\rho_s)$ leads to an increase of the flow $v_2$ with
bombarding energy. We point out that the increase of $v_2$ in PHSD
relative to HSD is also partly due to the higher interaction rates
in the partonic medium because of a lower ratio of $\eta/s$ for partonic
degrees of freedom at energy densities above the critical energy
density than for hadronic media below the critical energy 
density~\cite{Mattiello,Bass}. The relative increase in $v_3$ and $v_4$ in
PHSD essentially is due to the higher partonic interaction rate and,
thus, to a lower ratio $\eta/s$ in the partonic medium, which is
mandatory to convert initial spacial anisotropies to final
anisotropies in momentum space~\cite{Pet4}.

{\it Conclusions.}\
The anisotropic flows -- elliptic $v_2$, triangular $v_3$, and
hexadecupole $v_4$ -- are reasonably described within the PHSD model
in the whole transient energy range naturally connecting the
hadronic processes at moderate bombarding energies with
ultrarelativistic collisions at RHIC energies where the quark-gluon
DOF become dominant due to a growing number of partons. The
smooth growth of the elliptic flow with collision energy
demonstrates the increasing importance of partonic DOF This
feature is reproduced by neither explicit hadronic kinetic models
like HSD or UrQMD nor the AMPT model treating the partonic phase on
the basis of pQCD with massless partons and a noninteracting
equation-of-state for the partons. Further signatures of the
transverse collective flow, the higher-order harmonics of the
transverse anisotropy $v_3$ and $v_4$ change only weakly from
$\sqrt{s_{NN}}\approx$ 7~GeV to the top RHIC energy $\sqrt{s_{NN}}=$
200~GeV, roughly in agreement with preliminary experimental data.
Certainly, new measurements within the BES program at
RHIC, especially for higher-order harmonics, will further constrain
the partonic dynamics.

{\it Acknowledgements.}\
We are thankful to S.~Voloshin for constructive remarks and O.~Linnyk
for useful discussions. This work has been supported in part by the
DFG Grant WA 431/8-1, the DFG Grant CA 124/7-1, the RFFI Grants
08-02-01003-a, the Ukrainian-RFFI Grant 09-02-90423-ukr-f-a, and the
LOEWE center HIC for FAIR.

\end{document}